\documentclass[onecolumn,prl,unsortedaddress,superscriptaddress,floatfix,citeautoscript,nofootinbib,longbibliography]{revtex4-2}
\usepackage{adjustbox}
\usepackage{dcolumn}
\usepackage{latexsym,epsfig,graphicx,dcolumn,subfigure,comment,ulem}
\usepackage{amsmath,eqnarray,amssymb,amsbsy}
\usepackage{xcolor,color}
\usepackage[colorlinks,urlcolor=blue,citecolor=blue]{hyperref}

\begin{document}
\title{Non-Hermitian superconducting diode effect}

\author{Junjie Qi}
\email{qijj@baqis.ac.cn}
\affiliation{Beijing Academy of Quantum Information Sciences, 
	Beijing 100193, China}
 \author{Ming Lu}
 \affiliation{Beijing Academy of Quantum Information Sciences, 
	Beijing 100193, China}
\author{Jie Liu}
\affiliation{School of Physics, Xi’an Jiaotong University, Ministry of Education Key Laboratory for Non-Equilibrium Synthesis
and Modulation of Condensed Matter, Xi’an 710049, China}
 \author{Chui-Zhen Chen}
 \email{czchen@suda.edu.cn}
 \affiliation{School of Physical Science and Technology, Soochow University, Suzhou 215006, China}
\affiliation{
Institute for Advanced Study, Soochow University, Suzhou 215006, China}
\author{X. C. Xie}
\affiliation{International Center for Quantum Materials, School of Physics, Peking University, Beijing 100871, China}
\affiliation{
Interdisciplinary Center for Theoretical Physics and Information Sciences, Fudan University, Shanghai 200433, China}
\affiliation{Hefei National Laboratory, Hefei 230088, China}

\begin{abstract}  
The study of non-reciprocal phenomena has long captivated interest in both Hermitian and non-Hermitian systems.
The superconducting diode effect (SDE) is a non-reciprocal phenomenon characterized by unequal critical charge supercurrents flowing in opposite directions in Hermitian superconducting systems.
In this study, we introduce an SDE driven by non-Hermiticity in a superconducting quantum interference device (SQUID) under an external magnetic flux, which we refer to as the non-Hermitian SDE. 
Non-Hermiticity is introduced by coupling one of the two Josephson junctions to a gapless electron reservoir, introducing phase decoherence.
Remarkably, we find that an emergent non-Hermitian Fermi-Dirac distribution can give rise to SDE in the non-Hermitian SQUID.
We analyze the behavior of the SDE under both direct current (dc) and alternating current (ac) biases, highlighting the appearance of direction-dependent critical currents and asymmetric Shapiro steps as hallmarks of the SDE. 
Our findings not only reveal an experimentally accessible mechanism for non-Hermitian SDE but also open new avenues for investigating non-reciprocal phenomena in non-Hermitian systems.

\end{abstract}

\maketitle

\textit{Introduction.} The superconducting diode effect (SDE) is a non-reciprocal phenomenon that has recently emerged as a significant focus of research \cite{SDE1,SDE2,SDE3,SDE4,SDE5}.
The SDE enables supercurrent to preferentially flow in one direction while being suppressed in the opposite \cite{Ando,Wu}, 
offering the potential for applications in rectifiers, superconducting logic circuits, and directionally selective quantum sensors.
This phenomenon has been realized in both Josephson junctions (JJs) commonly referred to as the Josephson diode effect \cite{Chil,Baum,Jeon,Go,Wu,Baur,Pal,Vri,Diez,Turi,Chen,Shen,Trah,Mat}, and junction-free superconductors \cite{Ando,Ita,Miya,Kawa,Schu,Nar,Masu,Lin,Scam,Du,Cui}.
Achieving the SDE in JJs requires breaking both time-reversal and inversion symmetries \cite{SDE4}, typically via external magnetic fields and spin-orbit coupling \cite{CZC,Za,Bru,Yoko,Buz,Dol,Rey}. Alternative approaches include utilizing magnetic dopants \cite{YFS}, altermagnet \cite{QC1} or a chiral quantum dot (QD) \cite{QC2}. More complex configurations, such as multi-terminal JJs \cite{Cor}, asymmetric superconducting quantum interference devices (SQUIDs) \cite{Souto,Fom} and geometry control of the JJs \cite{Golod2019,Gerdemann1995,Maiellaro2024,Guarcello2024,Chesca2017},  further demonstrate the feasibility of the SDE. To date, all these mechanisms remain within the framework of Hermitian physics.

Meanwhile, non-reciprocity is also a key feature of non-Hermitian systems. Non-Hermitian physics has recently emerged as a rapidly developing field \cite{Bender,Moiseyev,Ganainy,Ashida,Bergholtz,Okuma}, offering a wealth of novel phenomena that are absent in its Hermitian counterparts. Significant progress has been made in non-Hermitian band theory, including the exploration of exceptional points \cite{Berry2004,Heiss2012,Zhen2015,Heiss2004,Dembowski,Doppler}, non-Hermitian skin effects \cite{ML,Zhang2022,Lin2023}, breakdown of conventional bulk-boundary correspondence \cite{Lee2016, Yao2018, Xiong2018,Lee2019, Yao2018a, Kunst2018, Yokomizo2019}, and extended topological classifications \cite{Shen2018,Gong2018, Kawabata2019, Kawabata2019a, Torres2019, Ghatak2019}. Despite these advances, 
exploring non-Hermitian physics in condensed matter experiments remains a formidable challenge, primarily due to the difficulty of precisely controlling gain and loss. Notably, the field of non-Hermitian quantum transport for condensed matter system has largely remained unexplored until recent developments \cite{Avila2019, Huang2020, Lin2022, Liu2022, Shao2023, Geng2023, Lu2024, Cayao2024a, Yan2024}. Previous studies have shown that non-reciprocal hopping terms can drive non-reciprocal physical processes \cite{Ochkan2024}. Remarkably, recent studies have further found that the non-Hermitian Fermi-Dirac distribution plays a significant role in governing the supercurrent in non-Hermitian JJs \cite{Cayao2023, Beenakker, PXS, Li2024, Kornich2023, Ohnmacht2024, Cayao2024b}. This discovery raises an intriguing question: can this distinctive non-Hermitian Fermi-Dirac distribution be leveraged to realize the SDE?
 \begin{figure}[ht!]
\centering
\includegraphics[width=0.5\linewidth]{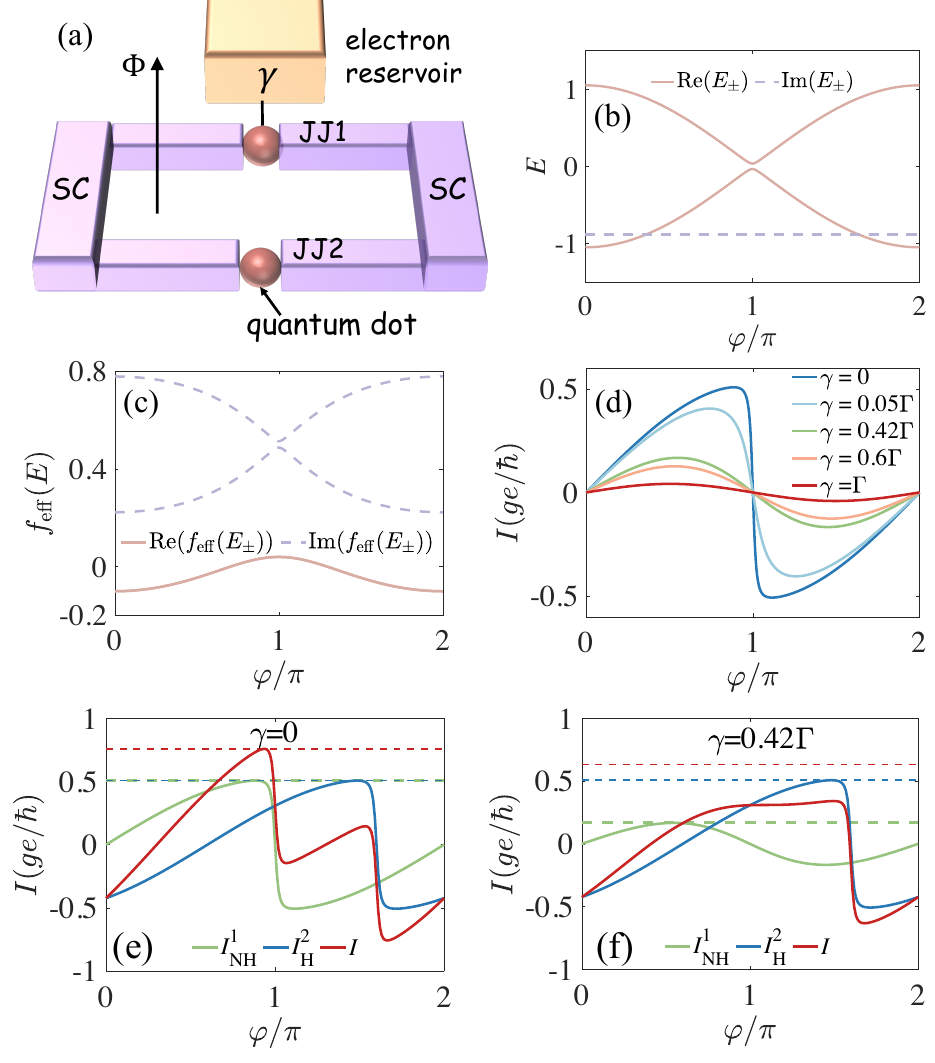}
\caption{(a) Schematic diagram of a non-Hermitian SQUID system, consisting of two superconductor-QD-superconductor
Josephson junctions, where non-Hermiticity is introduced to JJ1 by additionally coupling to a gapless electron reservoir with strength $\gamma$. 
The real and imaginary components of (b) the Andreev bound state spectrum $E_{\pm}$ and (c) the non-Hermitian Fermi-Dirac distribution $f_{\mathrm{eff}}(E_{\pm})$ as functions of $\varphi$ for $\gamma=0.42\Gamma$. (d)  Evolution of the CPR in a single junction, transitioning from the Hermitian ($\gamma = 0$) to the non-Hermitian regime ($\gamma = 0.05\Gamma$ to $\Gamma$).  CPRs for JJ1 ($I^{1}_\text{NH}$), JJ2 ($I^{2}_\text{H}$), and SQUID ($I$) with (e) $\gamma=0$ and (f) $\gamma=0.42\Gamma$. In (e) and (f), dashed horizontal lines refer to $I = I_{c-}$ and $\Phi=0.3\Phi_0$. Parameters: $\Gamma_L=1$, $\Gamma_R=1.1$, and $\epsilon_d=0$  \cite{SM1}}.
\label{fig:1}
\end{figure}

 In this study, we propose an experimental accessible non-Hermitian SDE in a non-Hermitian SQUID under an external magnetic flux. 
 Non-Hermiticity is introduced by coupling one of the JJs (JJ1) to a gapless electron reservoir (Fig.~\ref{fig:1}(a)), which induces phase decoherence. 
 We note that an asymmetric SQUID has been experimentally fabricated to explore the Hermitian SDE \cite{Ciaccia2022,ZhangPo,Greco2023,Greco2024,Paolucci,Yu2024}. Here, JJ1 exhibits complex Andreev bound state energies, and non-Hermitian Fermi-Dirac distributions,  resulting in a reduced supercurrent compared to the Hermitian regime. We analyze the supercurrent of the non-Hermitian SQUID under both direct current (dc) and alternating current (ac) biases. For dc bias, direction-dependent critical currents confirm the non-Hermitian SDE, with a maximum diode efficiency exceeding 37\%. Under ac bias, asymmetric Shapiro steps further demonstrate the presence of the non-Hermitian SDE. These findings unveil a novel mechanism for non-Hermitian SDE, emphasizing the role of the non-Hermitian Fermi-Dirac distribution in directional supercurrent transport.  Our work bridges non-Hermitian physics and superconducting phenomena, offering pathways for designing advanced superconducting devices with tunable directional transport properties.

\textit{Supercurrent in the non-Hermitian regime.} Our setup consists of an asymmetric SQUID formed by two JJs, denoted as JJ1 and JJ2, arranged as shown in Fig.~\ref{fig:1}(a). Each junction features a QD coupled to two $s$-wave superconducting electrodes. A distinctive feature of this setup is that JJ1 is coupled to a gapless electron reservoir with coupling strength $\gamma$,    which induces non-Hermitian effects by facilitating phase decoherence in the supercurrent flow \cite{Buttiker}. Quasiparticles are injected into the reservoir at a rate $1/\tau=2\gamma/\hbar$, which is assumed to be spatially uniform across the entire junction region. This process can be described by coupling the junction to a reservoir through a spatially extended tunnel barrier \cite{Zir1993, Beenakker1997}. Experimentally, such coupling can be realized in a JJ through a superconductor-insulator-normal metal (SIN) junction \cite{Gordeeva2020}.  In the weak-coupling, near-resonant regime, where $\epsilon_d, \Gamma \ll \Delta_0$, a single JJ coupled to a gapless electron reservoir can be described by an effective non-Hermitian Hamiltonian. This Hamiltonian can be solved analytically as \cite{Meng,Recher,Orie,Beenakker}
\begin{equation}\label{eq0}
\begin{aligned}
&\mathcal{H}_{\rm{eff}}=\left(\begin{array}{cc}
\epsilon_{d}-i \gamma & \frac{1}{2} e^{i \varphi_L} \Gamma_L+\frac{1}{2} e^{i \varphi_R} \Gamma_R \\
\frac{1}{2} e^{-i \varphi_L} \Gamma_L+\frac{1}{2} e^{-i \varphi_R} \Gamma_R & -\epsilon_{d}-i \gamma
\end{array}\right).
\end{aligned}
\end{equation}
Here,  $\epsilon_d$ is the energy level of the QD, and $\Gamma_{L/R}$ represent the coupling strengths between the dot and the left (right) superconducting electrodes.  The superconducting electrodes are described by their superconducting pair potential $\Delta_0$ and phase difference $\varphi=\varphi_R-\varphi_L$. The parameter 
$\gamma$ controls the non-Hermitian contribution to the system.

The Andreev bound state spectra are equal to two complex solutions of Eq.~\ref{eq0}  as $E_{\pm}=-i \gamma \pm \epsilon_{\mathrm{A}}$, where $\epsilon_{\mathrm{A}}$  is the Andreev bound state energy for a closed system, given by $ \epsilon_{\mathrm{A}}=\Delta_{\mathrm{eff}} \sqrt{1-\tau_{\mathrm{BW}} \sin^2(\varphi / 2)}$, $\Delta_{\mathrm{eff}}=\sqrt{\epsilon_d^2 + \frac{1}{4} \Gamma^2}$, and $\Gamma=\Gamma_L+\Gamma_R$. The normal-state transmission probability through the QD at the Fermi level $(\epsilon_F=0)$ is described by the Breit-Wigner formula  \cite{Breit} as $\tau_{\mathrm{BW}}=\frac{\Gamma_L \Gamma_R}{\epsilon_d^2+\frac{1}{4} \Gamma^2}$. The complex spectra $E_{\pm}$ as functions of $\varphi$ are shown in Fig.~\ref{fig:1}(b).

The influence of thermal noise on the supercurrent and related characteristics of the Josephson junction lies beyond the scope of this work \cite{Guarcello1,Guarcello2,Guarcello3,Guarcello4}; thus, for simplicity, we restrict our analysis to the zero-temperature limit. The zero-temperature supercurrent associated to complex spectrum $E_{\pm}$ of $\mathcal{H}_{\text {eff }}$ is determined by 
$I_{\text{NH}}(\varphi)=-\frac{e}{\pi\hbar} \frac{\mathrm{~d}}{\mathrm{~d} \varphi} \operatorname{Im} \operatorname{Tr}\left(\mathcal{H}_{\text {eff }} \ln \mathcal{H}_{\text {eff }}\right) $ \cite{PXS}, leading to the supercurrent:
\begin{equation}\label{eq02}
I_{\text{NH}}(\varphi)=-\frac{ge}{\pi \hbar} \frac{d}{d \varphi} \sum_{s=\pm}\operatorname{Im}\left[E_{s} \ln E_{s}\right]
\end{equation} 
with $g$ accounting for degeneracies (e.g., spin, valley, etc.),  $\hbar$ the reduced Planck constant, and $e$ the electron charge.
Here, $\ln (E_{\pm})$ arises from the non-Hermitian Fermi-Dirac distribution \cite{PXS} given by $f_{\mathrm{eff}}(E_{\pm})=-(1 / \pi) \ln (E_{\pm})$.
The real and imaginary components of $f_{\mathrm{eff}}(E_{\pm})$ as functions of $\varphi$ for $\gamma=0.42\Gamma$ are shown in Fig.~\ref{fig:1}(c). This non-Hermitian Fermi-Dirac distribution $f_{\mathrm{eff}}(E_{\pm})$ leads to the current-phase relation (CPR) at $T=0$ K (see SM for details) given by:
\begin{equation}\label{eq1}
I_{\text{NH}} (\varphi ) =I_{\text{H}}(\varphi ) \times \left[ \frac{2}{\pi} \arctan \left(\frac{\epsilon_{\mathrm{A}} (\varphi )} {\gamma}\right) \right],
\end{equation} 
in agreement with that obtained from the scattering matrix theory \cite{Beenakker}. Here, $I_{\text{H}}(\varphi ) =-\frac{g e}{\hbar} \frac{d \epsilon_{\mathrm{A}}(\varphi )}{d \varphi}$ describes the zero-temperature supercurrent in Hermitian regime. 
The reduction $\frac{2}{\pi} \arctan \left(\frac{\epsilon_{\mathrm{A}} (\varphi )} {\gamma}\right)=\text{Im} \sum\limits_{s=\pm} f_{\mathrm{eff}}(E_{s})$ originates form the non-Hermitian Fermi-Dirac distribution $f_{\mathrm{eff}}(E_{\pm})$.
Coulomb blockade effects are neglected in this open system, as they do not significantly affect the dynamics. When $\gamma \rightarrow 0$, the system is in the Hermitian regime. In this case, $\text{Im}f_{\mathrm{eff}}(E)=f_{\mathrm{FD}}(\text{Re}(E))= \Theta(-\text{Re}(E))$ is the zero-temperature Fermi-Dirac distribution, since $f_{\mathrm{eff}}(E)=-\frac{1}{\pi} \ln |\text{Re}(E)|+\mathrm{i} \Theta(-\text{Re}(E))$. According to the Eq.~\ref{eq02}, this Hermitian Fermi-Dirac distribution $f_{\mathrm{FD}}(E_{\pm})$ leads to the CPR as $I_{\text{H}}(\varphi ) =-\frac{g e}{\hbar} \frac{d \epsilon_{\mathrm{A}}(\varphi )}{d \varphi}$.

The evolution of the CPR in a
single junction from the Hermitian to the non-Hermitian regime is shown in Fig.~\ref{fig:1}(d). As $\gamma$ increases, the supercurrent decreases because electrons entering the electron reservoir lose phase coherence, rendering them unable to contribute to the supercurrent.  This reduction is quantitatively characterized by the factor $\text{Im} \sum\limits_{s=\pm} f_{\mathrm{eff}}(E_{s})=\frac{2}{\pi} \arctan \left(\frac{\epsilon_{\mathrm{A}} (\varphi )} {\gamma}\right)$, according to Eq.~\ref{eq1}.

\begin{figure}[ht!]
\centering
\includegraphics[width=0.5\linewidth]{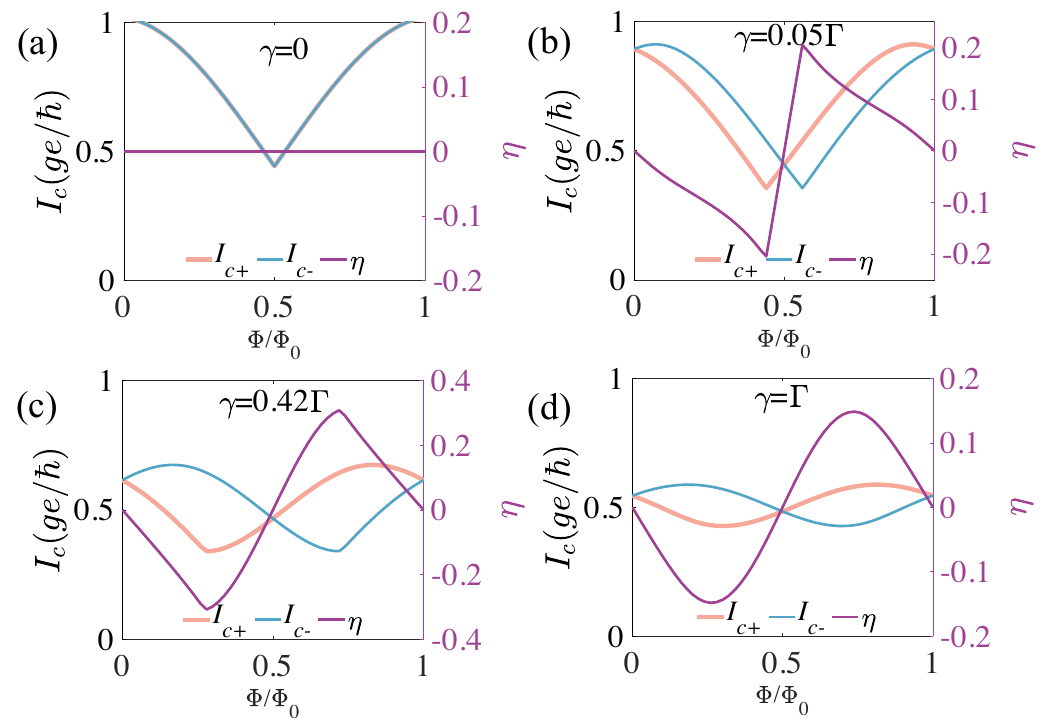}
\caption{(a)-(d) The critical Josephson currents of the SQUID, $I_{c\pm}$, exhibit direction-dependent values as a function of the magnetic flux $\Phi$, a hallmark of the SDE. The nonzero diode efficiency $\eta$ further demonstrates the presence of the SDE. In (a), $\gamma = 0$; In (b), $\gamma = 0.05 \Gamma$; In (c), $\gamma = 0.42 \Gamma$; In (d), $\gamma = \Gamma$. Parameters: $\Gamma_L=1$, $\Gamma_R=1.1$, $\epsilon_d=0$.}
\label{fig:2}
\end{figure}

\begin{figure}[ht!]
\centering
\includegraphics[width=0.5\linewidth]{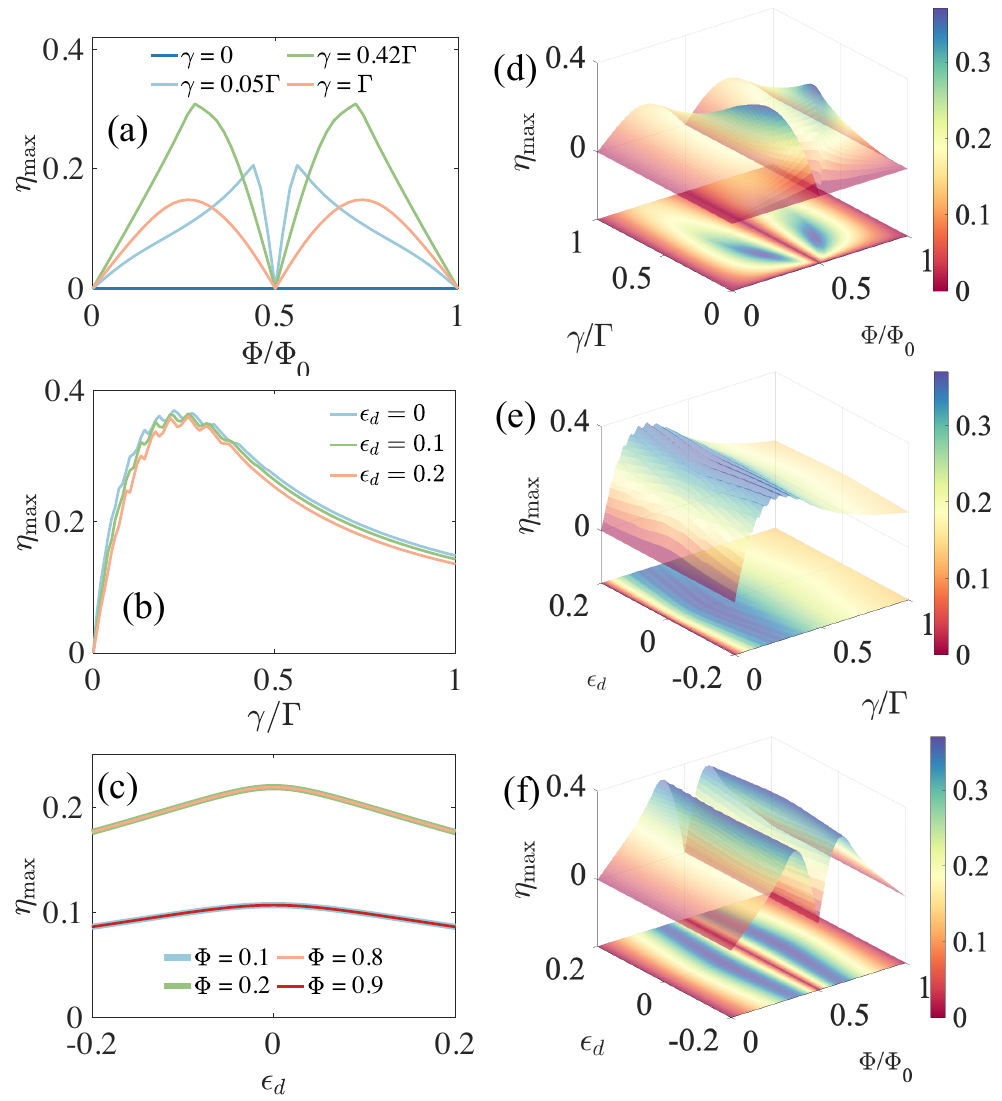}
\caption{The maximum diode efficiency $\eta_{\rm{max}}$ as a function of (a) the external magnetic flux $\Phi$ for various $\gamma$ values,  (b) the coupling strength $\gamma$ for various $\epsilon_d$ values,  and (c) the QD's energy level $\epsilon_d$ for various $\Phi$ values. In (c), the legend for $\Phi$ is expressed in units of $\Phi_0$, with $\Phi_0=1$. The maximum diode efficiency $\eta_{\rm{max}}$ as a bivariate function of : (d) $\Phi$ and $\gamma$, along with the projection of $\eta_{\rm{max}}$ on the $(\Phi,\gamma)$-plane; (e)  $\gamma$ and $\epsilon_d$, along with the projection of $\eta_{\rm{max}}$ on the $(\gamma, \epsilon_d)$-plane; and (f) $\Phi$ and $\epsilon_d$, along with the projection of $\eta_{\rm{max}}$ on the $(\Phi,\epsilon_d)$-plane. In (a) and (d), $\eta_{\rm{max}}=\rm{max}_{\epsilon_d}\left|\eta(\gamma,\Phi)\right|$; in (b) and (e), $\eta_{\rm{max}}=\rm{max}_{\Phi}\left|\eta(\gamma,\epsilon_d)\right|$; in (c) and (f), $\eta_{\rm{max}}=\rm{max}_{\gamma}\left|\eta(\Phi,\epsilon_d)\right|$. Parameters: $\Gamma_L=1$, $\Gamma_R=1.1$. }
\label{fig:3}
\end{figure}

\textit{dc-biased non-Hermitian SDE in the SQUID.} To introduce the SDE in an asymmetric SQUID, an external magnetic flux $\Phi$ is applied to break the time-reversal symmetry, as shown in Fig.~\ref{fig:1}(a). The asymmetry is tunable by coupling one of JJ1 to a gapless electron reservoir, introducing non-Hermitian dynamics into the system. This asymmetry in the SQUID's configuration is critical in realizing the diode effect, wherein the system allows a supercurrent to flow more easily in one direction than the other.

The total supercurrent flowing through the SQUID is the sum of the contributions from both junctions:
\begin{equation}\label{eq2}
I(\varphi)=I^{1}_{\text{NH}}(\varphi) +I^{2}_{\text{H}}(\varphi-2\pi\Phi/\Phi_0) 
\end{equation} 
where $\Phi_0=h/2e$ is the flux quantum, with the superscripts ``1" and ``2"  corresponding to JJ1 and JJ2, respectively. The individual supercurrents, $I^{1}_{\text{NH}}(\varphi)$ and $I^{2}_{\text{H}}(\varphi)$, are calculated using the CPR described in Eq.~\ref{eq1}. Self-inductance effects are neglected in this analysis.

We now investigate the non-Hermitian SDE in the SQUID. 
The CPRs for JJ1, JJ2, and the SQUID are shown in Figs.~\ref{fig:1}(e) and (f).  The dashed horizontal lines denote   $I=I_{c-}$, where $I_{c \pm}=\max[ \pm I(0 \leq \varphi<2\pi)]$ correspond to the positive and negative critical currents, respectively.
In Fig.~\ref{fig:1}(e), when $\gamma=0$,  the SQUID exhibits no SDE, as evidenced by the identical critical currents $I_{c+} = I_{c-}$. That's because $I(\varphi)=-I(2\pi\Phi/\Phi_0-\varphi)$ in Eq.~\ref{eq2} when $I^{1}_{\text{NH}}=I^{2}_{\text{H}}$ for $\gamma=0$, which leads to $I_{c+} = I_{c-}$. In contrast, for non-Hermitian case  ($\gamma \neq 0$), the coupling of JJ1 to a gapless electron reservoir introduces non-Hermitian effects such that $I^{1}_{\text{NH}}(\varphi)\neq I^{2}_{\text{H}}(\varphi)$. In this circumstance, the critical supercurrent $I_{c+} \neq I_{c-}$ as shown in Fig.~\ref{fig:1}(f), signals the non-Hermitian SDE in the SQUID. 
To quantify the non-reciprocity in the SQUID, we define the diode efficiency $\eta$ as:
\begin{equation}\label{eq3}
\eta  =\frac{I_{c+}-I_{c-}}{I_{c+}+I_{c-}}
\end{equation}
Figures \ref{fig:2}(a)-(d) show  $I_{c\pm}$ as functions of the magnetic flux $\Phi$ for $\gamma = 0$, $\gamma = 0.05 \Gamma$, $\gamma = 0.42 \Gamma$, and $\gamma = \Gamma$, respectively.  For the Hermitian case, SDE is absent with the diode efficiency $\eta=0$ for all values of $\Phi$ in Fig.~\ref{fig:2}(a). In the non-Hermitian SQUID, the diode efficiency $\eta\neq0$ in Figs.~\ref{fig:2}(b)-(d) confirms the presence of the SDE, except at specific flux values, $\Phi=n\Phi_0/2$ ($n \in \mathbb{Z}$), where the symmetry condition $I(\varphi)=-I(-\varphi)$ ensures $I_{c+} = I_{c-}$.

To better understand the maximum diode efficiency, Figs.~\ref{fig:3}(a)-(c) display $\eta_{\rm{max}}$ as a function of $\Phi$, $\gamma$, and $\epsilon_d$, respectively. For a given $\gamma$, the $\eta_{\rm{max}}$ curve is split into two symmetric parts around $\Phi = 0.5\Phi_0$, as they are time-reversal partners, as shown in Fig.~\ref{fig:3}(a).
This symmetry is further highlighted by the projection of $\eta_{\rm{max}}$ onto the $(\Phi, \gamma)$-plane in Fig.~\ref{fig:3}(d).
As a result, the analysis can be confined to the $(0, 0.5\Phi_0)$ range.
Notably, the maximum diode efficiency $\eta_{\rm{max}}$ in Fig.~\ref{fig:3}(a) increases as $\gamma$ rises from zero to a critical value $\gamma_c$, since a nonzero $\gamma$ enhances the asymmetry of the SQUID. However, beyond $\gamma_c$, the maximum diode efficiency $\eta_{\rm{max}}$ decreases with further increases in $\gamma$, due to a reduction in the supercurrent (see Fig.~\ref{fig:1}(d)). 
This nonmonotonic behavior of the trough values of $\eta_{\rm{max}}$ as a function of $\gamma$ is also evident in Fig.~\ref{fig:3}(b). 
In this case, the value of $\eta_{\rm{max}}$  exhibits an overall decreasing trend with local fluctuations  as the QD's energy level $\epsilon_d$ shifts from 0 to 0.2. Additionally, since $I(\varphi)$ is a function of $(\epsilon_d)^2$, $\eta_{\rm{max}}$ is symmetric about $\epsilon_d=0$, as shown in Fig.~\ref{fig:3}(c).  
Finally, to identify the highest efficiency $\eta_{\rm{max}}$ in our setup, we present $\eta_{\rm{max}}$ as  a bivariate function of  $\Phi$ and $\gamma$, $\gamma$ and $\epsilon_d$, as well as $\Phi$ and $\epsilon_d$ in Figs.~\ref{fig:3}(d)-(f), respectively.
It is found that $\eta_{\rm{max}}$ exceeding 37\% in all cases.

  
\begin{figure}[ht!]
\centering
\includegraphics[width=0.5\linewidth]{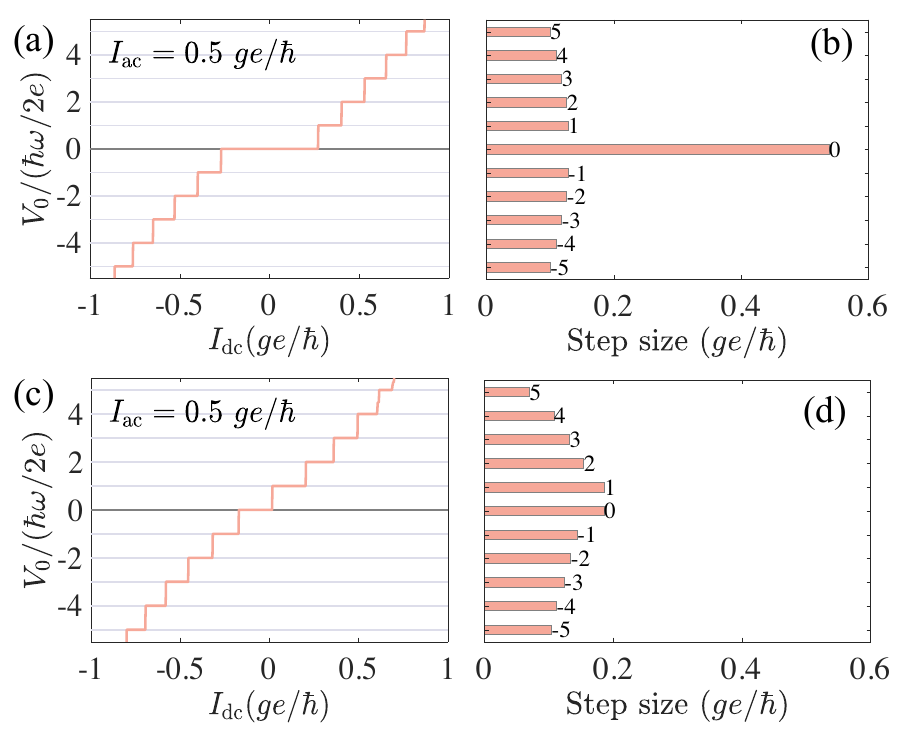}
\caption{(a) The current-voltage curve exhibits symmetric Shapiro steps, indicating the absence of ac-biased SDE. (b) Bar graph of the Shapiro step size from (a). (c) The current-voltage curve shows asymmetric Shapiro steps, signifying the presence of ac-biased SDE. (d) Bar graph of the Shapiro step size from (c).  In (a)-(d), $I_{\rm{ac}}= 0.5\ ge/\hbar$. In (c) and (d), $\gamma=0.42\Gamma$. Parameters: $\omega = 0.1 \times(2ge^2R/\hbar^2)$, $\Gamma_L=1$, $\Gamma_R=1.1$, $\epsilon_d=0$, $\Phi = 0.3\Phi_0$ \cite{Q}.}
\label{fig:4}
\end{figure}

\textit{Asymmetric Shapiro steps.} Next, we examine the characteristics of the non-Hermitian SDE under combined dc and ac bias conditions.  In our work, the system is assumed to be in the overdamped regime due to both the SQUID geometry formed by two SNS junctions and the non-Hermitian dissipation mechanism. Thus, this analysis is performed using the resistively shunted junction (RSJ) model \cite{RSJ1,RSJ2,RSJ3,Souto}, which describes Josephson dynamics by modeling the JJ as a parallel circuit comprising a JJ and a resistance $R$. The applied current has the form $ I_{\rm {dc}} + I_{\rm {ac}} \cos \omega t$. Within the RSJ framework, the total current through the SQUID is governed by a first-order differential equation describing the phase difference $\varphi$,
\begin{eqnarray}\label{eq4}
 &&I_{\rm {dc}} + I_{\rm {ac}} \cos \omega t = I[\varphi(t)] + \frac{\hbar}{2eR} \frac{d\varphi}{dt}
\end{eqnarray}
where $I_{\rm {dc}}$ and $I_{\rm {ac}}$ denote the amplitudes of the dc and ac current components, respectively, $\omega$ is the microwave frequency, and $R$ is the shunt resistance. The term $I[\varphi(t)]$ denotes the total SQUID supercurrent, arising from the contributions of both Josephson junctions JJ1 and JJ2, as expressed by Eq.~\ref{eq2}. 
Since the supercurrents in JJ1 and JJ2 are related by a magnetic-flux-induced phase shift (see Eq.\ref{eq2}), it is sufficient to consider only a first-order differential equation in the SQUID analysis.
Using the second Josephson equation, $V =\frac{\hbar}{2e}\frac{d \varphi}{d t}$, the dc voltage drop $V_0$ across the junction can be derived by solving Eq.~\ref{eq4} and computed as the time-averaged voltage $V_0=\left \langle V  \right \rangle_T$. The solution is computed numerically using the fourth-order Runge-Kutta method (see SM for details), with results detailed in Fig.~\ref{fig:4}. 




The SDE under ac bias is characterized by two key features, as illustrated in Fig.~\ref{fig:4}. (1) Offset of the zeroth Shapiro step: In a symmetric SQUID without SDE, the voltage exhibits discrete Shapiro steps \cite{Shapiro} at levels $V_0=n\hbar \omega/2e$, where $n=0,\pm1,\pm2,...$ (Fig. \ref{fig:4}(a)).  The zeroth Shapiro step is centered at $I_{\rm_{dc}}=0 $ on the dc current-voltage curve. In contrast, the presence of SDE causes a shift in the zeroth step, deviating from $I_{\rm_{dc}}=0 $, as depicted in Fig.~\ref{fig:4}(c). (2) Asymmetric Shapiro step: The bar graphs of Shapiro step sizes from Fig.~\ref{fig:4}(a) and (c) are shown in Fig.~\ref{fig:4}(b) and (d), respectively. In the absence of SDE, the Shapiro step sizes are symmetric about the zeroth step (Fig. ~\ref{fig:4}(b)).  However, with SDE, the symmetry is broken, resulting in unequal step sizes for the 
$n$-th and $-n$-th steps (Fig.~\ref{fig:4}(d)), reflecting the unidirectional nature of the supercurrent.  Differential resistance maps further corroborate these two features (see SM for details). These observations collectively confirm that ac bias induces SDE in our system. 

Furthermore, the presence of fractional Shapiro steps serves as evidence of a highly skewed CPR in the JJ, arising from  higher-harmonic contributions \cite{Raes2020,Ueda2020}. As shown in Fig.~\ref{fig:4}, the CPR at $\gamma = 0.42\Gamma$ remains approximately sinusoidal (see Fig.~\ref{fig:1}(d)), and consequently, fractional Shapiro steps are absent. By decreasing $\gamma$ to enhance the CPR skewness, fractional Shapiro steps emerge in our model (see SM for more details). 

 For the experimental realization of our model, nonadiabatic effects should be suppressed, as they can significantly modify the Shapiro steps \cite{Dubos2001,Basset2019}. In diffusive JJs, nonadiabatic effects become relevant when $\omega > 2E_{\rm{Th}}/h$, where $E_{\rm{Th}}$ is the Thouless energy. In ballistic JJs, the corresponding threshold is set by $\omega > I_{SW}R$, with $I_{SW}$ the switching current and $R$ the resistance \cite{Ueda2020}. Previous experimental studies have verified asymmetric Shapiro steps arising from several mechanisms \cite{Valentini2024,SHT2024}, thus the asymmetry driven by the non-Hermitian mechanism is also expected to be experimentally verified.

\textit{Discussion and conclusion.} In our work, the introduction of non-Hermiticity has twofold effects: First, non-Hermiticity breaks the inversion symmetry between JJ1 and JJ2, leading to an asymmetric SQUID. The observations above suggest that this asymmetry, induced by non-Hermiticity, enables the realization of a significant SDE with a maximum diode efficiency $\eta_{\rm{max}}$ exceeding 37\%. Second, non-Hermiticity significantly enhances diode efficiency. For instance, the unequal energy levels of QD in JJ1 and JJ2 can also induce an asymmetric SQUID in the Hermitian regime  ($\gamma=0$), resulting in the SDE. However, when non-Hermiticity ($\gamma \neq 0$) is introduced, it further increases the diode efficiency $\eta$ by 20\% (see SM for details).

In summary, our study reveals a novel mechanism for the SDE induced by the non-Hermitian Fermi-Dirac distribution in an asymmetric SQUID. The coupling of one JJ to a gapless electron reservoir generates phase decoherence, breaking inversion symmetry and enabling directional asymmetry in supercurrent flow. This effect manifests in both dc and ac biases, with hallmark observations including asymmetric critical currents and shifted Shapiro steps. Our work provides a deeper understanding of the interplay between non-Hermitian effects and superconducting phenomena, offering potential pathways for engineering advanced superconducting devices with tunable directional transport properties.

\textit{Acknowledgments.} We thank  Yu-Hang Li, Yue Mao and Xiao-Fang Shu for fruitful discussions. We are grateful to the support by the National Natural Science Foundation of China (Grant No.12204053), the National Key R\&D Program of China (Grants No.2022YFA1403700), and the Innovation Program for Quantum Science and Technology (Grant No.2021ZD0302400).
C.-Z. Chen is also support by the Natural Science Foundation of Jiangsu Province Grant (No.BK20230066)
and the Priority Academic Program Development (PAPD) of Jiangsu Higher Education Institution.

\textit{Data availability.} The data that support the findings of this article are openly available \cite{QJJ}.

\bibliographystyle{apsrev4-1}

\begin{thebibliography}{100}
\bibitem{Q}
When $\epsilon_d=0$ and $\Gamma_L=\Gamma_R$, $\epsilon_A$ can become zero for certain values of $\varphi$, leading to a divergence in the differential resistance. Therefore,  we set $\Gamma_L \neq \Gamma_R$ when $\epsilon_d=0$.
\end{thebibliography}

\clearpage

\section*{Supplemental Material}

\renewcommand{\thefigure}{S\arabic{figure}}
\setcounter{figure}{0}

\renewcommand{\theequation}{S\arabic{equation}}
\setcounter{equation}{0}

\section{I. Derivation of the supercurrent}

We begin by expressing the supercurrent in the non-Hermitian regime as:
\begin{equation}\label{eq1}
I_{\text{NH}}(\varphi)=-\frac{e}{\pi\hbar} \frac{\mathrm{~d}}{\mathrm{~d} \varphi} \operatorname{Im} \operatorname{Tr}\left(\mathcal{H}_{\text {eff }} \ln \mathcal{H}_{\text {eff }}\right)
\end{equation}
By introducing the eigenvalues $E_{\pm}$ of $\mathcal{H}_{\text {eff }}$ in Eq.~\ref{eq1}, the supercurrent can be written as:
\begin{equation}\label{eq2}
I_{\text{NH}}(\varphi)=-\frac{e}{\pi \hbar} \frac{d}{d \varphi} \operatorname{Im}\left[E_{+} \ln E_{+}+E_{-} \ln E_{-}\right]
\end{equation}
Using the expressions  $E_{\pm}=-i \gamma \pm \epsilon_{\mathrm{A}}$, the logarithm of a complex number \( E_\pm \) is given by $\ln E_\pm = \ln|E_\pm| + i \arg(E_\pm)$, where the modulus of $|E_\pm|$ is $ |E_\pm| = \sqrt{\epsilon_A^2 + \gamma^2}$, and $\arg(E_\pm)$ is the argument  of $E_\pm$ .

For $E_{+}$, since $\text{Re}(E_{+})=\epsilon_A>0$ and $\text{Im}(E_{+})=-\gamma<0$, the argument is
\begin{equation*}
\arg(E_{+}) = \arctan\left(\frac{-\gamma}{\epsilon_A}\right).  
\end{equation*}
 For $E_{-}$, since $\text{Re}(E_{-})=-\epsilon_A<0$ and $\text{Im}(E_{-})=-\gamma<0$, the argument is
\begin{equation*}
\arg(E_{-}) = \arctan\left(\frac{\gamma}{\epsilon_A}\right)-\pi.  
\end{equation*}
Thus, using the identity  $\arctan \left(\frac{1}{x}\right)=\frac{\pi}{2}-\arctan (x)$, the supercurrent is 

\begin{equation}\label{eq3}
I_{\text{NH}}(\varphi)=-\frac{e}{\pi \hbar} \frac{d}{d \varphi}\left[-2 \gamma \ln \sqrt{\epsilon_A^2+\gamma^2}+ 2 \epsilon_A \arctan \left(\frac{\epsilon_A}{\gamma}\right)\right]
\end{equation}
After simplification, the supercurrent becomes:
\begin{equation}\label{eq4}
I_{\text{NH}}(\varphi)=-\frac{e}{\pi \hbar} \left[2 \frac{d \epsilon_A}{d \varphi} \arctan \left(\frac{\epsilon_A}{\gamma}\right)\right].
\end{equation}

\section{II. The effects of $\tau_{\rm{BW}}$}

\begin{figure}[ht!]
\centering
\includegraphics[width=0.7\linewidth]{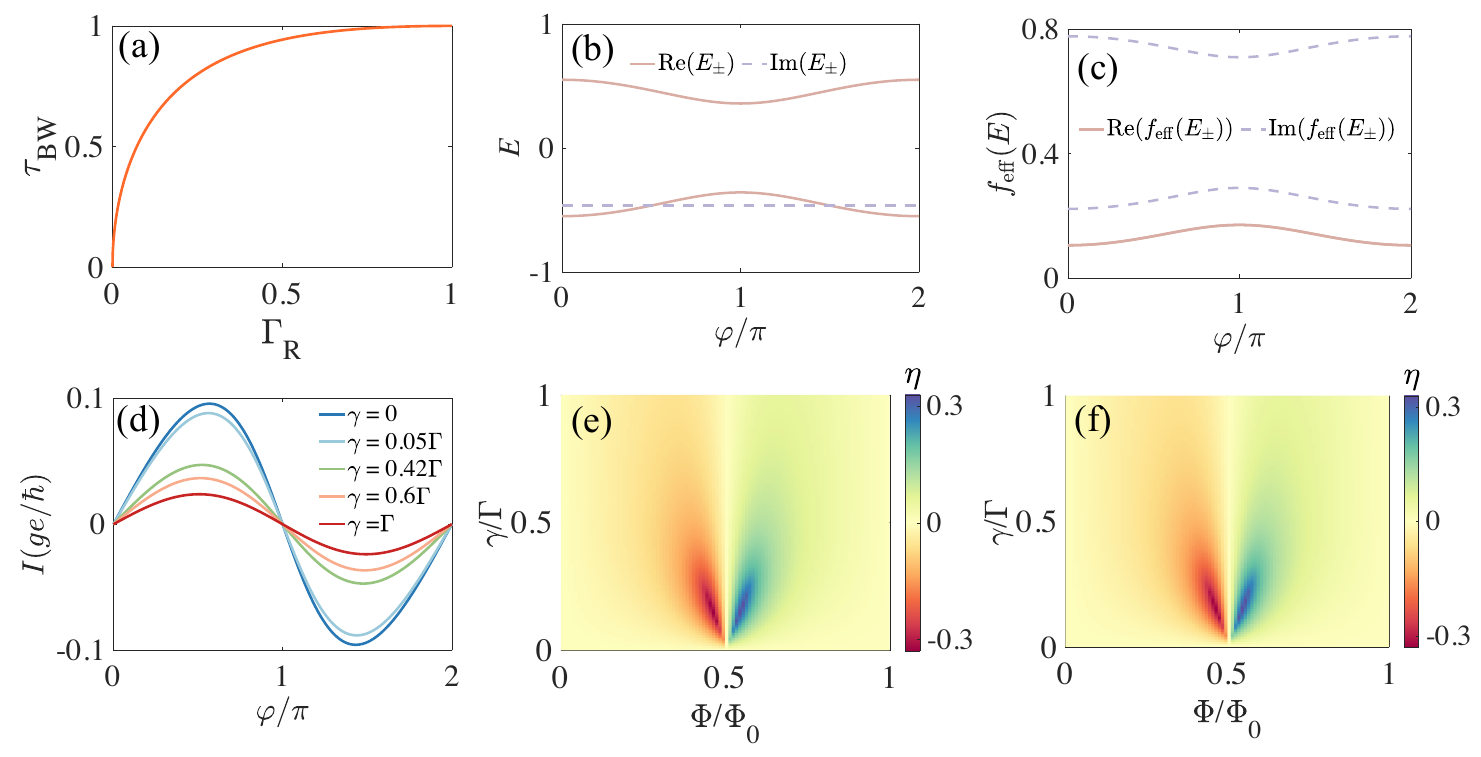}
\caption{ (a) The transparency $\tau_{\rm{BW}}$ as a function of $\Gamma_R$. 
The real and imaginary components of (b) the Andreev bound state spectrum $E_{\pm}$ and (c) the non-Hermitian Fermi-Dirac distribution $f_{\mathrm{eff}}(E_{\pm})$ as functions of $\varphi$ for $\gamma=0.42\Gamma$. (d)  Evolution of the current-phase relation in a single junction, transitioning from the Hermitian ($\gamma = 0$) to the non-Hermitian regime ($\gamma = 0.05\Gamma$ to $\Gamma$). In (b)-(d), $\Gamma_R=0.1$ and $\tau_{\rm{BW}}=0.575$.
The diode efficiency as functions of $\gamma$ and $\Phi$ when (e) $\Gamma_R=0.1$ and $\tau_{\rm{BW}}=0.575$ and (f) $\Gamma_R=0.2$ and $\tau_{\rm{BW}}=0.7454$. Parameters: $\Gamma_L=1$.}
\label{fig:S1}
\end{figure}

In this section, we examine the effects of $\tau_{\rm{BW}}$, which varies with the coupling strength $\Gamma_R$. As shown in Fig.~\ref{fig:S1}(a), $\tau_{\rm{BW}}$ increases monotonously with $\Gamma_R$. Figures~\ref{fig:S1}(b) and (c) show the real and imaginary components of the Andreev bound state spectrum $E_{\pm}$ and  the non-Hermitian Fermi-Dirac distribution $f_{\mathrm{eff}}(E_{\pm})$ as functions of $\varphi$ for $\gamma=0.42\Gamma$, respectively. When $\Gamma_R=0.1$ and $\tau_{\rm{BW}}=0.575$, the current-phase relation approximates a sine function for $\gamma=0$, as depicted in Fig.~\ref{fig:S1}(d). When $\gamma \neq 0$, the supercurrent is reduced due to the non-Hermitian Fermi-Dirac distribution. Finally, Figs.~\ref{fig:S1}(e) and (f), demonstrate that the diode efficiency exceeds 30\% even at lower values of $\tau_{\rm{BW}}$ compared to those presented in the main text.

\section{III. The enhanced diode efficiency by the non-Hermiticity}

\begin{figure}[ht!]
\centering
\includegraphics[width=0.5\linewidth]{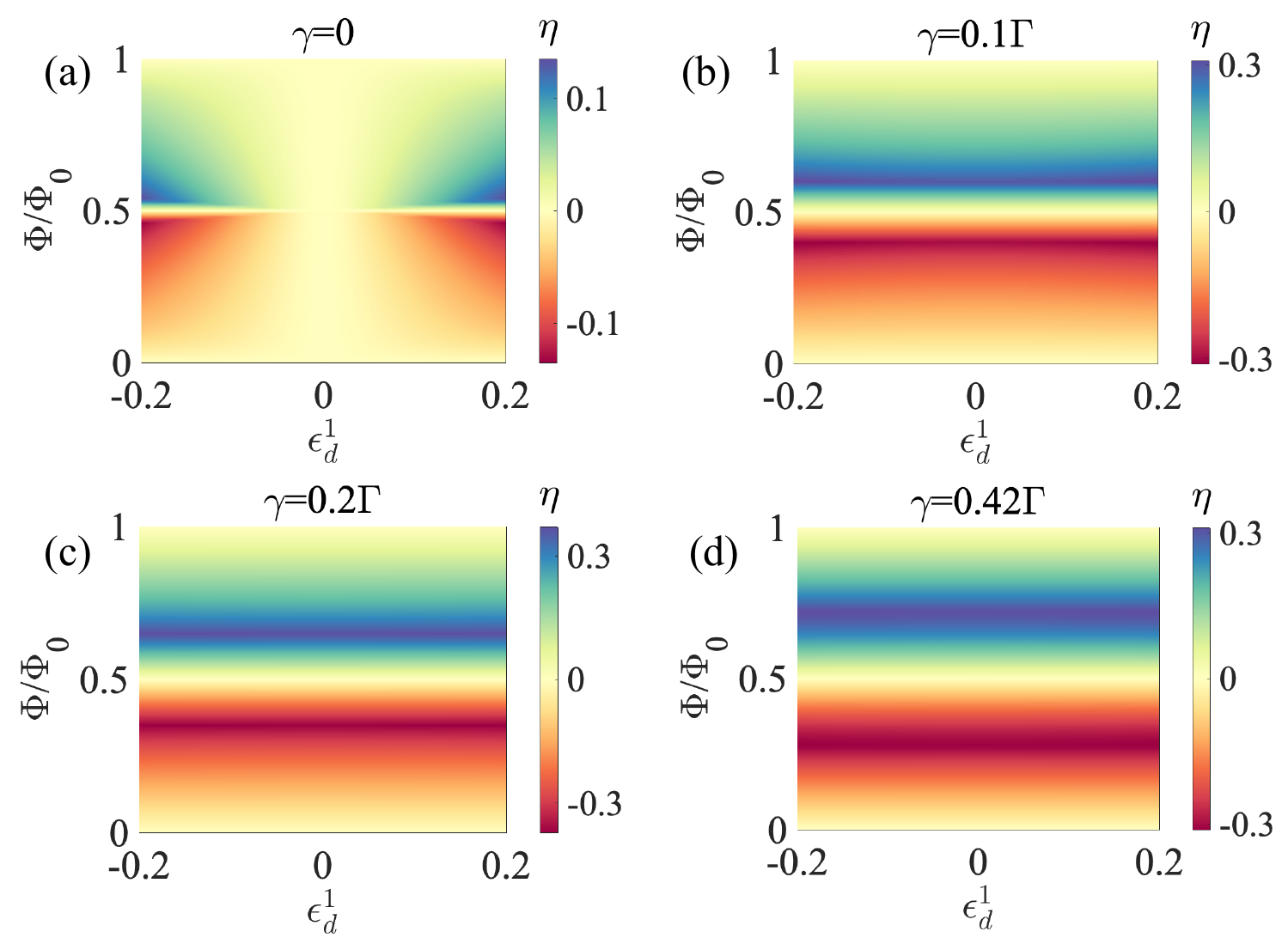}
\caption{ The diode efficiency as functions of the QD's energy level in JJ1 ($\epsilon_d^1$) and the external magnetic flux ($\Phi$) for various $\gamma$ values, with the QD's energy level ($\epsilon_d^2$) in JJ2 set to zero.  Parameters: $\Gamma_L=1$, $\Gamma_R=1.1$.}
\label{fig:S2}
\end{figure}

The unequal energy levels of QD in JJ1 and JJ2 induce an asymmetric SQUID in the Hermitian regime  ($\gamma=0$), leading to the SDE with a diode efficiency $\eta$ greater than 10\%, as illustrated in Fig.~\ref{fig:S2}(a).  However, non-Hermiticity significantly enhances $\eta$, offering a further advantage of our setup. As shown in Figs.~\ref{fig:S2}(b)-(d), $\eta$ exceeds 30\% for all cases.

\section{IV. The procedure of the fourth order Runge-Kutta method}

The detailed procedure for applying the fourth-order Runge-Kutta method to solve Eq.~6 is outlined below. Define the function

\begin{equation}\label{eq1}
f(t,\varphi) =\dot \varphi= \frac{2eR}{\hbar}(I_{\rm {dc}}+I_{\rm {ac}} \cos w t-I[\varphi(t)]),
\end{equation}
Divide the time interval $t$ into discrete steps of size $h$. At each time step $t_n$ $(n=0,1,2,...,)$, calculate $\varphi_{n+1}$ using:
\begin{equation}\label{eq2}
\begin{aligned}
\varphi_{n+1} &= \varphi_{n} + \frac{h}{6}(k_1+2k_2+3k_3+k_4),\\
t_{n+1} &= t_{n}+h
\end{aligned}
\end{equation}
where the intermediate coefficients are:
\begin{equation*}\label{eq3}
\begin{aligned}
k_1 &= f(t_n,\varphi_n),\\
k_2 &= f(t_n+\frac{h}{2},\varphi_n+h\frac{k_1}{2}),\\
k_3 &= f(t_n+\frac{h}{2},\varphi_n+h\frac{k_2}{2}),\\
k_4 &= f(t_n+h,\varphi_n+hk_3),\\
\end{aligned}
\end{equation*}
The initial conditions are set as $t_0$=0 and $\varphi_n(0)=0$.

\section{V. Differential resistance maps}

\begin{figure}[ht!]
\centering
\includegraphics[width=0.5\linewidth]{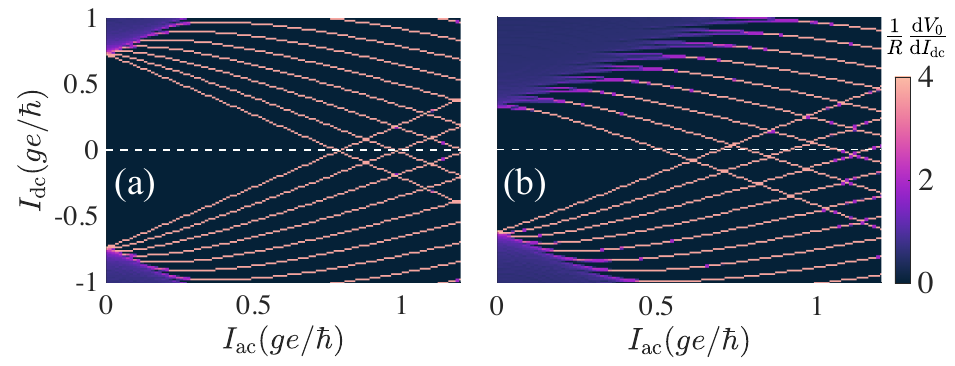}
\caption{Differential resistance maps, $dV_0/dI_{\rm{dc}}$, are plotted as functions of $I_{\rm{dc}}$ and $I_{\rm{ac}}$ in (a) (without SDE) and (b) (with SDE).  In (b), $\gamma=0.42\Gamma$. Parameters: $\omega = 0.1 \times(2ge^2R/\hbar^2)$, $\Gamma_L=1$, $\Gamma_R=1.1$, $\epsilon_d=0$, $\Phi = 0.3\Phi_0$.}
\label{fig:S3}
\end{figure}

 The differential resistance $dV_0/dI_{\rm{dc}}$ as functions of $I_{\rm{dc}}$ and $I_{\rm{ac}}$ is shown in Fig.~\ref{fig:S3}. In the absence of SDE, both the differential resistance values and their separations are symmetric about $I_{\rm_{dc}}=0$ for Fig.~\ref{fig:S3}(a). In contrast, under SDE (Fig.~\ref{fig:S3}(b)), the zeroth step shifts from $I_{\rm_{dc}}=0 $ (white dash line), and the separations of adjacent differential resistances become asymmetric for a fixed $I_{\rm{ac}}$. This asymmetry is consistent with the unequal Shapiro step sizes observed in Fig.~4(d) in the main text.

\section{VI. fractional Shapiro steps}

\begin{figure}[ht!]
\centering
\includegraphics[width=0.7\linewidth]{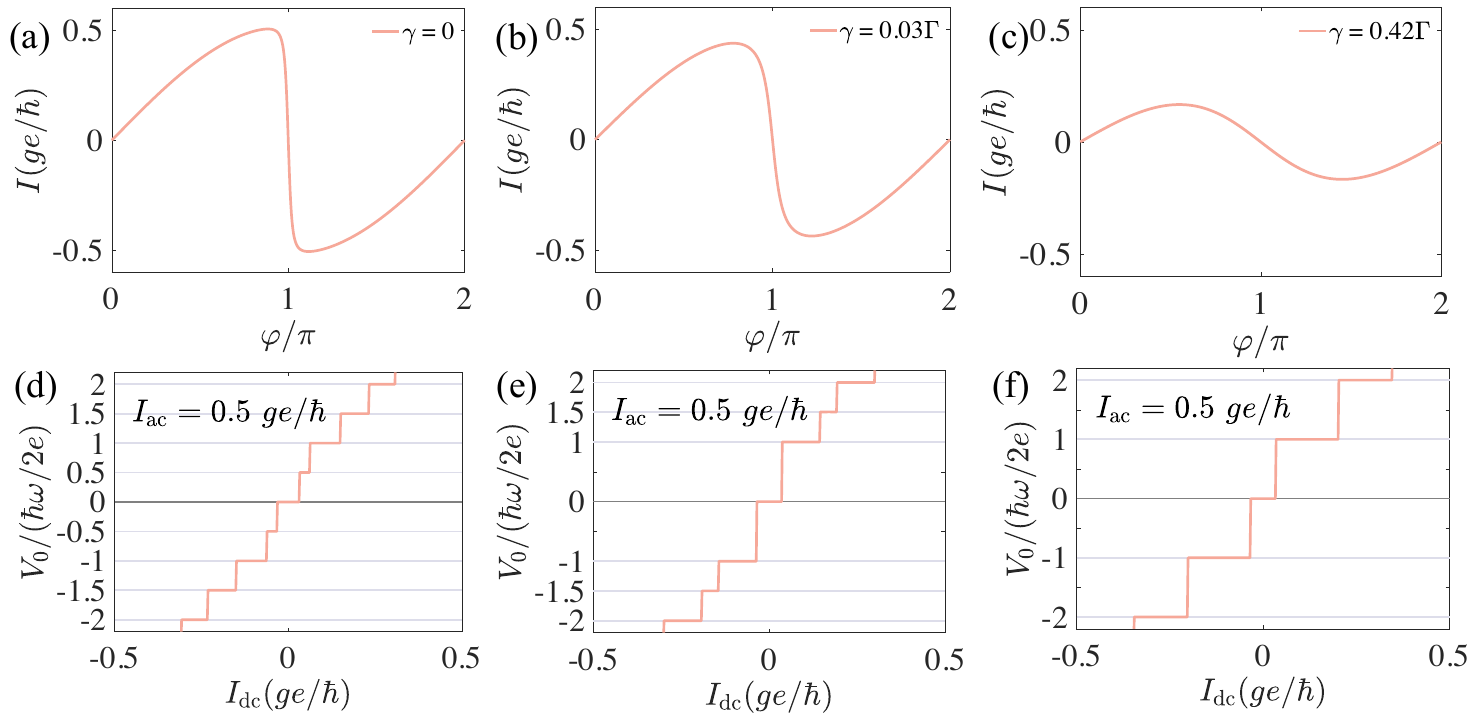}
\caption{(a)-(c) The current-phase relations for a single junction.  (d)-(f) The current-voltage curves exhibiting Shapiro steps. In (a) and (d), $\gamma=0$; In (b) and (d), $\gamma=0.03\Gamma$; In (c) and (f), $\gamma=0.42\Gamma$. Parameters: $\Phi = 0.5\Phi_0$. All other parameters are identical to those used in Fig. 4 of the main text.}
\label{fig:S4}
\end{figure}

In our work, fractional Shapiro steps are also observed in the regime where the CPR of a single junction is highly skewed, consistent with previous studies, as shown in Fig.~\ref{fig:S4}. For $\gamma=0$ in Fig.~\ref{fig:S4} (a) and (d), the CPR remains strongly skewed, and fractional Shapiro steps are clearly visible.  As the $\gamma$ value increases, the CPR gradually evolves from a skewed to an approximately sinusoidal form, leading to the progressive disappearance of the fractional steps. In the main text, we set $\gamma=0.42\Gamma$ (same as Fig.~\ref{fig:S4} (c) and (f)), where the CPR of the SQUID is not skewed enough and fractional Shapiro steps are absent.

\end{document}